\documentclass[sigconf]{acmart}
  
\usepackage{graphics, epstopdf, epsfig}
\usepackage{amstext}
\usepackage{subfigure}
\usepackage{color}
\usepackage{indentfirst}
\usepackage{multirow}
\usepackage{amsmath}
\usepackage{amssymb}
\usepackage{multicol}
\usepackage{rotating}
\usepackage{booktabs}
\usepackage{algorithm}
\usepackage{algpseudocode}
\usepackage{fixltx2e}
\usepackage{xcolor}
\usepackage{lipsum,graphicx}
\usepackage{balance}

\def\BibTeX{{\rm B\kern-.05em{\sc i\kern-.025em b}\kern-.08emT\kern-.1667em\lower.7ex\hbox{E}\kern-.125emX}}
    
\copyrightyear{2019}
\acmYear{2019}
\setcopyright{acmcopyright}
\acmConference[CIKM '19] {The 28th ACM International Conference on Information and Knowledge Management}{November 3--7, 2019}{Beijing, China}
\acmBooktitle{The 28th ACM International Conference on Information and Knowledge Management (CIKM'19), November 3--7, 2019, Beijing, China}
\acmPrice{15.00}
\acmDOI{10.1145/3357384.3357892}
\acmISBN{978-1-4503-6976-3/19/11}

\begin{document}

\fancyhead{}

\title{DBRec: Dual-Bridging Recommendation via Discovering Latent Groups}

\author{Jingwei Ma}
\affiliation{%
  \institution{The University of Queensland}}
\email{jingwei.ma@uq.edu.au}

\author{Jiahui Wen}
\authornote{Corresponding author}
\affiliation{%
  \institution{National University of Defense Technology}}
\email{wen\_jiahui@outlook.com}

\author{Mingyang Zhong}
\affiliation{%
  \institution{Central Queensland University; Chongqing Meiqi Industry Co.}}
\email{m.zhong@cqu.edu.au}

\author{Liangchen Liu}
\affiliation{%
 \institution{The University of Queensland}}
\email{l.liu9@uq.edu.au}
 
\author{Chaojie Li}
\affiliation{%
  \institution{Aliexpress, Alibaba group}
  \city{Hang Zhou}
  \state{Zhe Jiang}
  \postcode{311121}}
\email{Cjlee.cqu@163.com}

\author{Weitong Chen}
\affiliation{%
  \institution{The University of Queensland}}
\email{w.chen9@uq.edu.au}

\author{Yin Yang}
\affiliation{\institution{Hamad Bin Khalifa University}}
\email{yyang@hbku.edu.qa}

\author{Hongkui Tu}
\affiliation{\institution{National University of Defense Technology}}
\email{tuhkjet@foxmail.com}

\author{Xue Li}
\affiliation{\institution{The University of Queensland}}
\email{xueli@itee.uq.edu.au}
%
\renewcommand{\shortauthors}{Trovato and Tobin, et al.}

%
\begin{abstract}
In recommender systems, the user-item interaction data is usually sparse and not sufficient for learning comprehensive user/item representations for recommendation. To address this problem, we propose a novel dual-bridging recommendation model (DBRec). DBRec performs latent user/item group discovery simultaneously with collaborative filtering, and interacts group information with users/items for bridging similar users/items. Therefore, a user's preference over an unobserved item, in DBRec, can be bridged by the users within the same group who have rated the item, or the user-rated items that share the same group with the unobserved item. In addition, we propose to jointly learn user-user group (item-item group) hierarchies, so that we can effectively discover latent groups and learn compact user/item representations. We jointly integrate collaborative filtering, latent group discovering and hierarchical modelling into a unified framework, so that all the model parameters can be learned toward the optimization of the objective function. We validate the effectiveness of the proposed model with two real datasets, and demonstrate its advantage over the state-of-the-art recommendation models with extensive experiments.
\end{abstract}

%
%


%

%
\maketitle

\section{Introduction}
Due to the prevalence of the Internet, recommender systems have attracted remarkable attentions in industrial and research communities. Recommender systems mainly aim to alleviate the problem of information overload for online services, and explore a tremendous amount of information and recommend interested items for each user. Collaborative Filtering (CF) is one of the most popular recommendation techniques, which leverages users' historical behaviors to collaboratively infer user preferences over items. Among the CF-based methods, Matrix Factorization (MF) is efficient in capturing user preferences and providing superior recommendation performance. The basic idea behinds matrix factorization is to embed each user/item into a low-dimensional vector, and model a user-item rating with the interaction (i.e. inner product) between the corresponding user/item vector. 

Many previous works have developed MF-based methods for modeling user-item interactions. For example, neural matrix factorization \cite{he2017neural} combines generalize matrix factorization and multiple-layer perceptron for learning the interaction function. The multiple-layer of nonlinear transformations is proved to be efficient in learning user-item interactions, and yields state-of-the-art recommendation performance. ConvNCF \cite{he2018outer} utilizes outer product to transform each user-item embedding pair into a two-dimensional interaction map, and employs convolution and pooling layers for modeling high-order user-item correlations. \cite{cheng2018delf} involves context items (e.g. items rated by users) for modeling user-item interactions, hence user preferences over items are comprised of user-item and item-item similarities. Eventhough the aforementioned methods are shown to be effective, they suffer from the problem of data sparseness, as they are mainly based on user-item interactions, and users usually give few ratings to the items in real scenarios.

To address the data sparseness problem, existing research propose to incorporate additional information for boosting recommendation performance. Some of them \cite{wang2017item, yang2017bridging} leverage social networks for bridging similar users, and propagating user preferences along the social links. Some others resort to attribute data of different modalities to mitigate data sparseness. Typical attribute data include user profile \cite{ma2018lga}, descriptive texts \cite{cheng2019mmalfm, cheng2016effective, cheng2018aspect}, images \cite{liu2018multi, liu2017best}, etc. However, these recommender models are tightly coupled with one information source or another, and the requirement of extra information sources limits their scalability. 

In this work, we propose a novel method for boosting recommendation without the requirement of auxiliary data sources. To this end, we discover latent user/item groups and model user-item interactions collaboratively. The discovered latent groups are then interacted with users/items to boost recommendation performance. The rationale underlying DBRec is that, the estimation of a user's rating over an unobserved item, in the proposed model, can be bridged by the users within the same group who have interacted with the item, or the user-rated items that share the same group with the unobserved item. The in-group users and items are abstracted with high-level group representations, and the interactions of user-item group and item-user group result in a dual-bridging architecture. In this light, the dual-bridging mechanism can mitigate data sparseness through bridging similar users/items. Also, we utilize deep neural networks for modeling user-user group (item-item group) hierarchies, so that users/items are forced to be close to their groups for learning compact representations. 

Notice that, in \cite{ma2019multi}, it exploits group information to boost social recommendation. However, the group information is explicitly required, hence those methods cannot be scaled to the scenarios where the group information is not explicitly available. 

The contributions of this work can be concluded as follows, 

\begin{itemize}
\item We propose to seamlessly combine collaborative filtering and latent group discovery for recommendation. The discovered group information can dual-bridge similar users and items for boosting recommendation performance.

\item We propose to jointly model user-user group (item-item group) hierarchies for learning compact user/item representations. 

\item We validate the effectiveness of the proposed model on two real datasets, and demonstrate its advantage over the state-of-the-art recommendation models with comparison experiments and comprehensive analysis. 
\end{itemize}

\section{Related Work}
In this section, we review the representative state-of-the-art CF-based recommender models. 
\subsection{Neural Recommendation}
In recently years, deep learning techniques have been widely applied to recommender community. The majority of works employ multiple-layer neural networks to model deep user-item interactions. For example, NeuMF \cite{he2017neural} combines generalized matrix factorization and multiple-layer perceptron for modeling user-item similarities. ConvMF \cite{he2018outer} proposes to use outer product to transform embeddings of each user-item pair into a two-dimensional interaction map, and then employs convolutional and pooling layer to model high-order interrelations among embedding dimensions. Cheng et al.\shortcite{cheng2018delf} leverage item contexts (i.e. historical items rated by users) to compensate the interaction function. Ebesu et al. \shortcite{ebesu2018collaborative} propose an attention to aggregate neighboring users, and jointly exploit the neighborhoods with user-item interactions to derive the recommendation. However, one major shortcoming of those models is that, they are mainly based on the interaction data, and suffer from data sparseness. 

\subsection{Recommendation with Side Information}
To alleviate data sparseness and boost recommendation performance, existing works incorporate side information of different modalities for recommendation. The auxiliary information reflects item characteristics, and can bridge similar items (users) for alleviating data sparseness. In most cases, they employ deep neural networks to extract high-level representations for the extra information, and populate them into CF-based methods for recommendation. The extra data sources that are commonly exploited include user profile \cite{ma2018lga}, texts \cite{cheng2019mmalfm}, images \cite{liu2018multi} and demands \cite{li2017data, li2018integrating}. However, the aforementioned methods are usually coupled with one type of information source or another, and lack the scalability when they are deployed for recommendation in different scenarios. Notice that group information has been exploited for group recommendation. For example, in \cite{ma2019multi}, the authors address group recommendation problem, and aggregate representations of group members with an attention mechanism to obtain group representation. However, the group information needs to be specified in the data, and hence those methods are inapplicable in our case.

In \cite{ebesu2018collaborative}, the authors combine latent factor model and neighborhood-based structure, and propose an attention mechanism to find similar users based on the specific user and item. However, they simply aggregate similar users for estimating the rating scores, therefore the neighboring users are not sufficiently leveraged for bridging unobserved user-item pairs. Chen et al. \cite{chen2019social} aggregate social friends with an attention mechanism given the specific items, and interact the aggregated neighborhoods with the item to derive the recommendation. However, the interactions between the neighborhoods and the target items exert too strict regularization between the users and their social friends.

\subsection{Community Detection}
The proposed latent group discovery is related to community detection \cite{cavallari2017learning} that identifies groups of densely connected nodes in a graph. However, those methods \cite{kozdoba2015community} explicitly require proximity information among the nodes for discovering latent communities, and they are inapplicable in the case when the link information is not available. 

By contrast, we propose to discover latent user/item groups and model user-item interactions collaboratively, the proposed latent group discovery and collaborative filtering can be mutually beneficial to each other. On one hand, users/items with similar rating activities are more likely to have similar representations in collaborative filtering, which facilitates the user/item clustering process. On the other hand, user/item groups can bridge similar users/items and boost recommendation performance. 

\section{The proposed model}
\subsection{Preliminaries}
In a recommendation problem, we have a set of $m$ user $U = \{u_1,\cdots,u_m\}$, a set of $n$ items $V=\{v_1,\cdots,v_n\}$ and a rating matrix $\mathbf{R}\in\{0,1\}^{m\times n}$, where each element $r_{ij}\in\{0,1\}$ equals 1 if user $u_i$ has an observed interaction (e.g. clicked, viewed, etc.) with item $v_j$ and 0 otherwise. The task of item recommendation is to estimate the rating score $\hat{r}_{ij}$ for unobserved user-item interactions, and recommend for each user the items with high estimated rating scores. In this paper, we also propose to discover latent user groups $G_u = \{1,...,k\}$ and item groups $G_v = \{1,...,k\}$ for bridging similar users and items, where $k$ is the pre-specified number of groups for users and items. 

In this paper, two-dimension matrices are denoted with bold uppercase symbols, single-dimension vectors are denoted bold lowercase symbols, and scalars are represented with lowercase symbols. The symbols used in this paper are summarized in Table.\ref{tb:symbol}, other intermediate symbols are described in the corresponding sections. 
\begin{table}
\small
\caption{Descriptions of the symbols used in this paper}
\label{tb:symbol}
\begin{tabular}{c|p{5.5cm}}
\hline
Symbols & Descriptions \\
\hline
$m$,$n$ & number of users and items\\
$u_i$,$v_j$ & $i$-th user and $j$-th item\\
$U,V$ & set of users and items\\
$\mathbf{R}$ & rating matrix\\
$r_{ij}$ & rating score of $u_i$ over $v_j$\\
$\hat{r}_{ij}$ & estimated rating score of $u_i$ over $v_j$\\
$\mathbf{u}_i,\mathbf{v}_j$ & embedding of $u_i$ and $v_j$\\
$d$ & dimension of user and item embeddings\\
\hline
$G_u$,$G_v$ & set of user and item groups\\
$\mathbf{G}_u, \mathbf{G}_v$ & embedding matrix of user and item groups\\
$\mathbf{g}_{s}^u$ & embedding of $s$-th user group\\
$\mathbf{g}_{t}^v$ & embedding of $t$-th item group\\
$a_i$ & group label of user $u_i$\\
$b_j$ & group label of item $v_j$\\
$d_g$ & dimension of user and item group embeddings\\
\hline
$\boldsymbol{\mu}_i$ & soft latent group representation of $u_i$\\
$\mathbf{u}_i'$& reconstruction of $\mathbf{u}_i$\\
$\mathbf{v}_j'$& reconstruction of $\mathbf{v}_j$\\
$\{\mathbf{u}_{i_s}|s=1,...,p\}$ & $p$ negative samples for $\mathbf{u}_i'$ in learning user groups\\
$\{\mathbf{v}_{j_t}|t=1,...,p\}$ & $p$ negative samples for $\mathbf{v}_j'$ in learning item groups\\
\hline
$\phi_{\{1,...,L\}}^{\{uv,ug,vg,u,v\}}$ & L layers neural network for modeling user-item, user-group, item-group, user and item hierarchies\\
\hline
\end{tabular}
\end{table}

\subsection{Basic model}
Inspired by previous works \cite{he2017neural, manotumruksa2017deep}, we employ multiple neural networks as the basic model for capturing deep user-item interactions. Without loss of generality, given the embedding vectors $\mathbf{u}_i\in\mathbb{R}^{d}, \mathbf{v}_j\in\mathbb{R}^{d}$ for user $u_i$ and item $v_j$ respectively, where $d$ is the number of dimensions of user and item embedding spaces. The corresponding rating score of $u_i$ over $v_j$ can be estimated as:
\begin{equation}
\begin{split}
\mathbf{z}_0 &= [\mathbf{u}_i;\mathbf{v}_j;\mathbf{u}_i\circ\mathbf{v}_j]\\
\mathbf{z}_{ij} &= \phi_L(...\phi_1(\mathbf{z}_0) ...)\\
\phi_l(\mathbf{z}_{l-1}) &= \sigma_l(\mathbf{W}_l\mathbf{z}_{l-1}+\mathbf{b}_l),l\in[1,L]\\
\hat{r}_{ij} &= \frac{1}{1+exp(-\mathbf{w}_{uv}^T\mathbf{z}_{ij})}
\end{split}
\end{equation}
where $[;]$ and $\circ$ are the concatenation and element-wise multiply operation respectively. $L$ is the number of hidden layers in the neural network, and $\mathbf{W}_l, \mathbf{b}_l, \sigma_l$ are the weight matrix, bias vector and non-linear activation function of the $l$-th layer, respectively. $\mathbf{w}_{uv}$ is the parameter vector for transforming high-level user-item interaction vector $\mathbf{z}_{ij}$ into a logit, which is then used for rating score estimation. 

\subsection{Latent group discovery}

In many online services, some users (items) have the same interests (characteristics) and constitute a group. Users/items in a group share the same group representation, and it can bridge similar users/items and mitigate data sparseness. However, in most cases, the group information is not explicitly available. In this paper, we propose to discover latent user/item groups and model user-item interactions collaboratively. Specifically, we propose to learn user and item group embeddings, and introduce embedding matrix $\mathbf{G}_u\in\mathbb{R}^{k\times d_g}$ and $\mathbf{G}_v\in\mathbb{R}^{k\times d_g}$ for user and item group respectively, where $k$ is the number of user/item group as specified previously, and it is much smaller than $m$ and $n$. $d_g$ is the dimension of user/item group embeddings, and it is assumed to be smaller than the user/item dimension $d$ as group embeddings represent users/items at a more general level of granularity. 

\begin{figure}
\centering
\includegraphics[width=5cm]{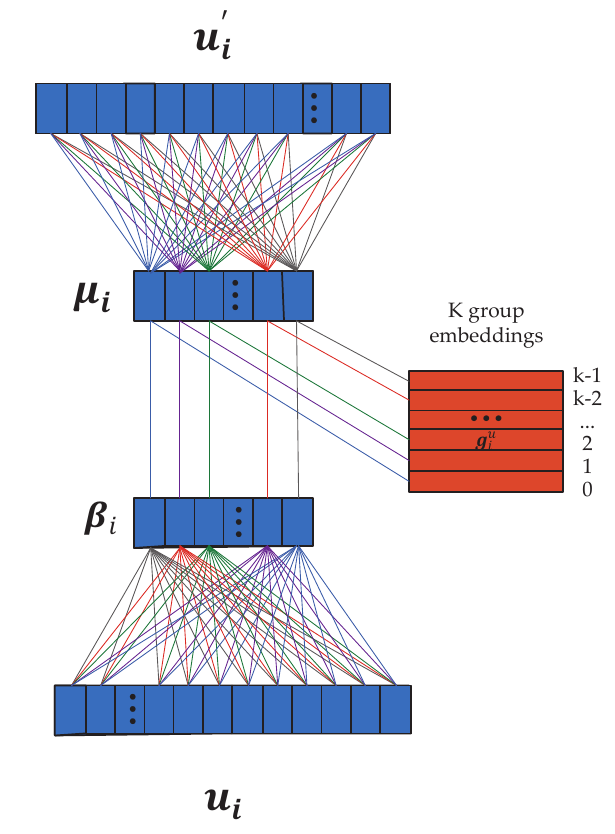}
\caption{Illustration of user group learning process.}
\label{fig:lgd}
\end{figure}

For learning user group embeddings, as an example, each user embedding is input into a non-linear transformation to obtain a vector that has the dimension of $k$, which is equal to the number of user groups. The value of each dimension of the vector can be viewed as the weight that the user belongs to the corresponding group, and the weighted sum of the user group embedding matrix is regarded as the soft latent group representation of the user. The user embedding is then reconstructed from the soft latent group representation. Therefore, user group embeddings are learned by transforming user embeddings into their reconstructions with least possible of information loss, and preserving most of the information in the $k$ embedded user groups. In addition, users with similar representations are supposed to have similar soft latent group representations, and are clustered automatically into the same user group. The user group learning process is illustrated in Fig.\ref{fig:lgd}, and we detail the learning process in the following subsections.

\subsubsection{Soft latent group representation}
Each user embedding $\mathbf{u}_i$ is fed into a non-linear transformation to obtain a vector $\boldsymbol{\beta}_i\in\mathbb{R}^k$, as shown in the following equation. 

\begin{equation}
\boldsymbol{\beta}_i = softmax(\mathbf{W}_u\mathbf{u}_i+\mathbf{b_u})
\end{equation}
where $\mathbf{W}_u\in\mathbb{R}^{k\times d}$ is the weighted matrix parameter, and $\mathbf{b_u}\in\mathbf{R}^k$ is the bias vector. $\boldsymbol{\beta}_i$ can be viewed as the activations toward the user group embeddings, and each dimension of $\boldsymbol{\beta}_i$ is the probability that the user belongs to the related group. The weighted sum of the user group embeddings with respect to $\boldsymbol{\beta}_i$ is then regarded as the soft latent group representation of user $u_i$. 

\begin{equation}
\label{eq:slgr}
\boldsymbol{\mu}_i = \sum_{s=1}^k\beta_{i,s} \mathbf{g}_s^u
\end{equation}
where $\beta_{i,s}$ is the $s$-th dimension of vector $\boldsymbol{\beta}_i$, and $\mathbf{g}_s^u$ is the $s$-th row of embedding matrix $\mathbf{G}_u$, representing the embedding of $s$-th user group. $\boldsymbol{\mu}_i\in\mathbb{R}^{d_g}$ is the soft latent group representation of user $u_i$. Therefore, users with similar embeddings are more likely to have similar activations toward the group embeddings, and they are automatically clustered into the same group in an unsupervised manner. 

\subsubsection{Reconstruction user embeddings}
We have obtained the soft latent group representations for the users. Now we describe how to calculate the reconstruction of the embedding for each user $u_i$ from his/her soft latent group representations $\boldsymbol{\mu}_i$. We reconstruct user embeddings from their related soft latent group representations $\boldsymbol{\mu}_i$. The underlying reason is that if two users have similar soft latent group representations, then they are more likely within the same user group and the reconstructions of their embeddings are assume to be similar. In this paper, we employ non-linear transformations to reconstruct user embeddings, as shown follows:
\begin{equation}
\label{eq:recst}
\mathbf{u}_i'=sigmoid(\mathbf{W_{u'}}\boldsymbol{\mu}_i+\mathbf{b}_{u'})
\end{equation}
where $\mathbf{W_{u'}}\in\mathbb{R}^{d\times d_g}$ is the weighted matrix parameter, and $\mathbf{b}_{u'}\in\mathbb{R}^d$ is the bias vector. At this point, the reconstruction of user embeddings is similar to an autoencoder, and obtaining the soft latent group representations can be viewed as the encoder stage of an autoencoder that takes user embeddings as input and maps them to the latent space while preserving group information. Eq.(\ref{eq:recst}) is similar to the decoder stage of an autoencoder that maps the soft latent group representation to the reconstruction $\mathbf{u}_i'$ of the same shape as $\mathbf{u}_i$, as illustrated in Fig.\ref{fig:lgd}.

\subsubsection{Learning objective}
Learning user groups is equivalent to minimizing the reconstruction error of $\mathbf{u}_i'$ from $\mathbf{u}_i$. We employ the contrastive max-margin objective function that is commonly used in previous work \cite{mohit2016feuding, socher2014grounded}. Specifically, for each user embedding $\mathbf{u}_i$, we randomly sample $p$ users as negative users. The embeddings of the negative samples are represented as $\{\mathbf{u}_{i_s}|s=1,2,...,p\}$, and the learning objective is to make the reconstructed embedding $\mathbf{u}_i'$ similar to its origin embedding $\mathbf{u}_i$ while different from those negative embeddings. Therefore, the objective of user group learning $\mathcal{L}_u$ is defined as a hinge loss that maximize the cosine similarity between $\mathbf{u}_i'$ and $\mathbf{u}_i$ and simultaneously minimize that between $\mathbf{u}_i'$ and the negative samples:
\begin{equation}
\mathcal{L}_u = \sum_{u_i}\sum_{s=1}^p max(0,1-\mathbf{u}_i'^T\mathbf{u}_i+\mathbf{u}_i'^T\mathbf{u}_{i_s})
\end{equation}

Similarly, the objective of item group learning can be formulated as:
\begin{equation}
\mathcal{L}_v = \sum_{v_j}\sum_{t=1}^p max(0,1-\mathbf{v}_j'^T\mathbf{v}_j+\mathbf{v}_j'^T\mathbf{v}_{j_t})
\end{equation}
where $\mathbf{v}_j'$ is the reconstructed embedding from $\mathbf{v}_j$. The calculation of $\mathbf{v}_j'$ is similar to that of $\mathbf{u}_i'$ and we leave out the detail description here to avoid redundancy. $\{\mathbf{v}_{j_t}|t=1,2...,p\}$ are $p$ negative samples for $\mathbf{v}_j'$ when computing objective $\mathcal{L}_v$.

\subsection{Hierarchies modeling}
\begin{figure*}
\centering
\includegraphics[width=12cm]{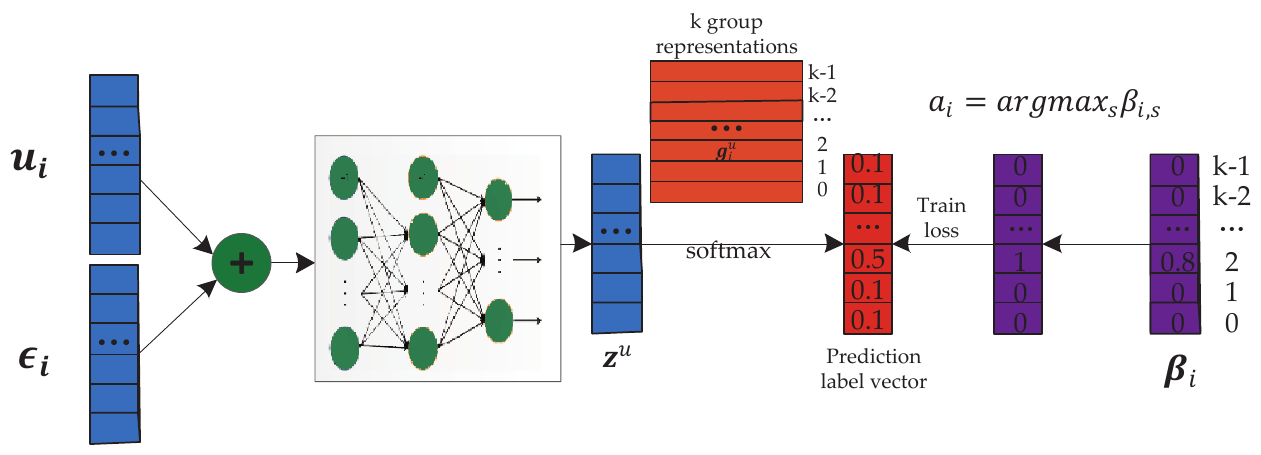}
\caption{Hierarchy modeling of user and user group.}
\label{fig:hierarchy}
\end{figure*}

The intuition of hierarchy modeling is that users/items should be close to their groups to learn compact representations. To this end, take user-user group hierarchy for example, we introduce a latent variable $\boldsymbol{\epsilon}_i$ for each user $u_i$, and minimize the distance between $\mathbf{u}_i+\boldsymbol{\epsilon}_i$ and its group representation $\boldsymbol{\mu}_i$ in the embedding space \cite{gao2018rec}. In this paper, we propose to utilize deep neural networks to model the user-user group (item-item group) hierarchies. 

We present the user-user group hierarchy modeling in Fig.\ref{fig:hierarchy}. As illustrated in the figure, we feed the vector $\mathbf{u}_i+\boldsymbol{\epsilon}_i$ into a multiple-layer neural network, and obtain a high-level representation $\mathbf{z}^{u_i}$ that has the same dimensions as the group representations. The similarities between $\mathbf{z}^{u_i}$ and all the group representations are measured with the inner product operation, and the resulted logits are transformed into a posterior probability vector with a softmax function, as shown in Eq.(\ref{eq:hierarchy}). 

\begin{equation}
\label{eq:hierarchy}
\begin{split}
\mathbf{z}_0^{u_i} &= \mathbf{u}_i+\boldsymbol{\epsilon}_i\\
\mathbf{z}^{u_i} &= \phi_L^u(...\phi_1^u(\mathbf{z}_0^{u_i})...)\\
a_i &= argmax_s(\beta_{i,s}),\quad s=1,2,...,k\\
\hat{r}^{u_i} &= \frac{exp({\mathbf{g}_{a_i}^u}^T\mathbf{z}^{u_i})}{\sum_{s=1}^kexp({\mathbf{g}_s^u}^T\mathbf{z}^{u_i})}
\end{split}
\end{equation}
where $\beta_{i,s}$ is described in Eq.(\ref{eq:slgr}), $a_i$ is the label of the user group that user $u_i$ has the maximum activation, and $\mathbf{g}_{a_i}^u$ is the corresponding group embedding, hence $\mathbf{g}_{a_i}^u$ can be viewed as the hard latent group representation of $u_i$. $\hat{r}^{u_i}$ is the posterior probability that user $u_i$ belongs to group $a_i$, and therefore bridging the gap between a user and his/her group is equivalent to maximizing the corresponding posterior probability. In other words, the neural network takes a user and his/her group representations as input, and produces a predicted distribution over the groups. The training objective is defined as the minimization between the predicted distribution and the distribution of actual labels \cite{oh2018know}, and it is demonstrated to be better than other loss functions \cite{shi2018proje}. The objective function of user-user group hierarchy can be defined as follows:
\begin{equation}
\mathcal{L}_{uu}=-\sum_{u_i}log\frac{exp({\mathbf{g}_{a_i}^u}^T\mathbf{z}^{u_i})}{\sum_{s=1}^kexp({\mathbf{g}_s^u}^T\mathbf{z}^{u_i})}
\end{equation}
Similarly, the objective function of item-item group hierarchy can defined as:
\begin{equation}
\mathcal{L}_{vv}=-\sum_{v_j}log\frac{exp({\mathbf{g}_{b_j}^v}^T\mathbf{z}^{v_j})}{\sum_{t=1}^kexp({\mathbf{g}_t^v}^T\mathbf{z}^{v_j})}
\end{equation}
where $b_j$ is group label of item $v_j$, and it is calculated in the same way as $a_i$.  $\mathbf{g}_{b_j}^v$ is the corresponding group embedding of $v_j$. The neural network for modeling item-item group hierarchy has the same architecture as that in Fig.\ref{fig:hierarchy}, and $\mathbf{z}^{v_j}$ can be viewed as the high-level representation of item $v_j$ extracted from the neural network.

\subsection{Dual-bridging collaborative filtering}
\begin{figure*}
\centering
\includegraphics[width=12cm]{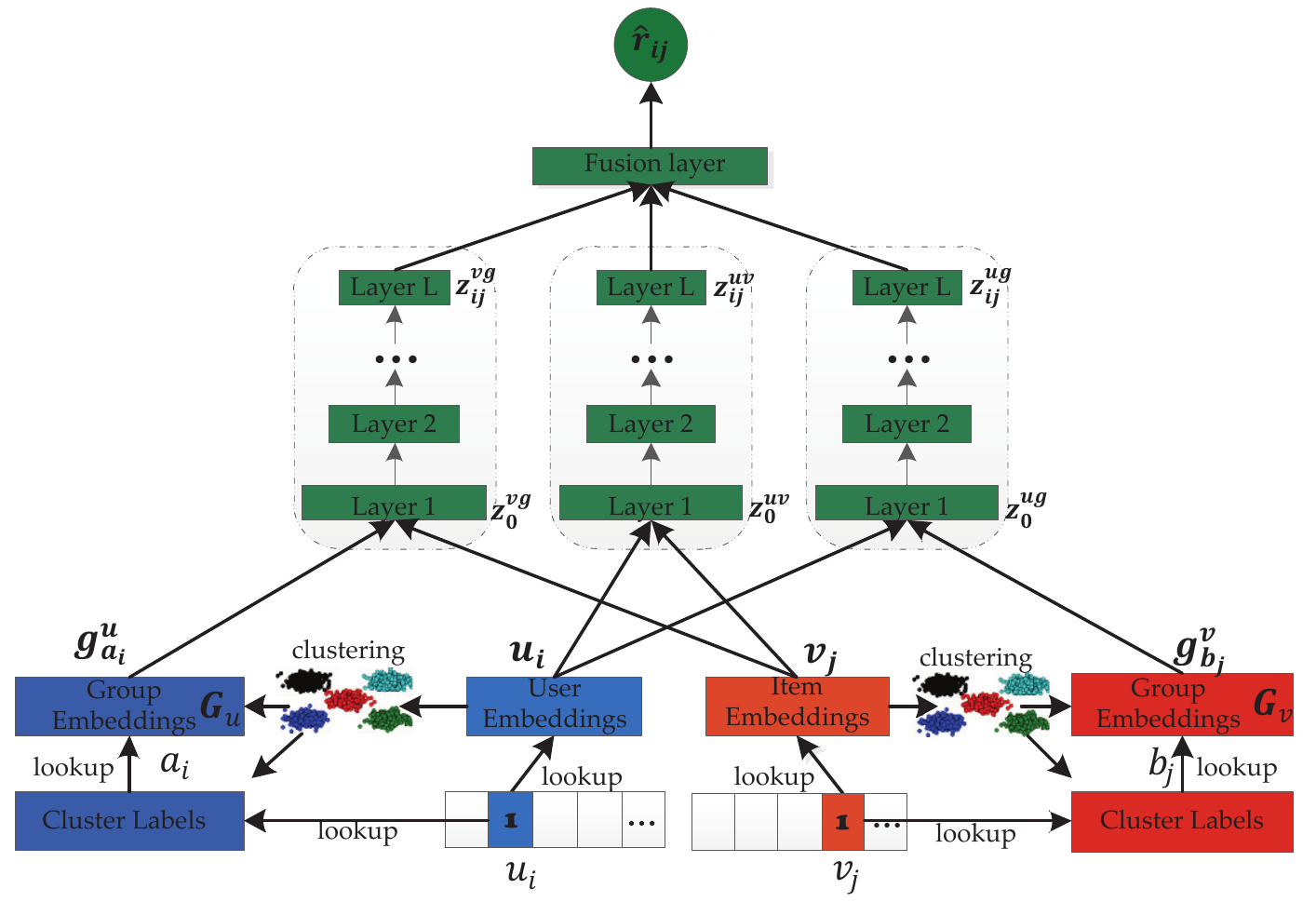}
\caption{Illustration of the dual-bridging collaborative filtering architecture.}
\label{fig:network}
\end{figure*}

Once we have obtained group representations, we can incorporate them for boosting recommendation performance. The underlying reason is that, the user-item group (item-user group) interactions can bridge similar users/items and alleviate data sparseness. For example, even though a user's interest in an item is unknown, the item can still be ranked higher in the recommendation list, as long as the user has positively interacted with some other items sharing the same group with the unobserved item, or users within the same group have positively rated the unobserved item. Notice that, the collaborative filtering and latent group discovery are iteratively performed, and they mutually benefit each other for better learning representations. First, collaborative filtering learns similar user/item representations based on the user-item interaction data, and this facilitates the clustering process for discovering users/items with similar representations. In turn, the latent group information can bridge similar users (items), and boost recommendation performance.

The general architecture of the proposed model is presented in Fig.\ref{fig:network}. As shown in the figure we introduce three neural networks for modeling user-item, user-item group, user group-item interactions respectively. The rating score of a user $u_i$ over an item $v_j$ is predicted as follows,
\begin{equation}
\label{eq:networks}
\begin{split}
\mathbf{z}_0^{uv} &= [\mathbf{u}_i;\mathbf{v}_j;\mathbf{u}_i\circ\mathbf{v}_j]\\
\mathbf{z}_{ij}^{uv} &= \phi_L^{uv}(...\phi_1^{uv}(\mathbf{z}_0^{uv}) ...)\\
\end{split}
\end{equation}

\begin{equation}
\begin{split}
\mathbf{z}_0^{ug} &= [\mathbf{u}_i;\mathbf{g}_{b_j}^v;\mathbf{u}_i^T\mathbf{M}_u\mathbf{g}_{b_j}^v]\\
\mathbf{z}_{ij}^{ug} &= \phi_L^{ug}(...\phi_1^{ug}(\mathbf{z}_0^{ug}) ...)\\
\end{split}
\end{equation}

\begin{equation}
\begin{split}
\mathbf{z}_0^{vg} &= [\mathbf{g}_{a_i}^u;\mathbf{v}_j; \mathbf{v}_j^T\mathbf{M}_v\mathbf{g}_{a_i}^u]\\
\mathbf{z}_{ij}^{vg} &= \phi_L^{vg}(...\phi_1^{vg}(\mathbf{z}_0^{vg}) ...)\\
\end{split}
\end{equation}

\begin{equation}
\label{eq:fuse}
\hat{r}_{ij}=\frac{1}{1+exp[-(\mathbf{w}_{uv}^T\mathbf{z}_{ij}^{uv}+\mathbf{w}_{ug}^T\mathbf{z}_{ij}^{ug}+\mathbf{w}_{vg}^T\mathbf{z}_{ij}^{vg})]}
\end{equation}
where $\mathbf{M}_u\in\mathbb{R}^{d\times d_g}$ is a transformation matrix for measuring the similarity between $\mathbf{u}$ and $\mathbf{g}_{b_j}^v$ in a common latent space, while $\mathbf{M}_v\in\mathbb{R}^{d\times d_g}$ is a matrix for measuring the similarity between $\mathbf{v}_j$ and $\mathbf{g}_{a_i}^u$. $\phi_{\{1,\cdots,L\}}^{\{uv,ug,vg\}}$ are the respective neural networks for modeling user-item, user-item group, item-user group interaction, and $\mathbf{w}_{\{uv,ug,vg\}}$ are the parameter vectors for transforming high-level representations (i.e. $\mathbf{z}_{ij}^{uv}$,$\mathbf{z}_{ij}^{ug}$,$\mathbf{z}_{ij}^{vg}$ ) of different interaction spaces into the final predictive logit. Therefore, with the dual-bridging architecture, an item can be ranked higher in the recommendation list for a user, as long as the user's preference over the item is properly modeled in any one of the three interaction spaces. The training objective of the proposed dual-bridging collaborative filtering can be defined as follows,
\begin{equation}
\mathcal{L}_{uv} = -\sum_{(u_i,v_j)\in\mathcal{D}}r_{ij}log\hat{r}_{ij}+(1-r_{ij})log(1-\hat{r}_{ij})
\end{equation}
where $\mathcal{D}$ is the training set.

\subsection{Model learning}
Due to the nature of implicit feedback and the large number of items in the recommendation task \cite{hu2018conet}, negative sampling \cite{mikolov2013distributed} is employed to approximate $\mathcal{L}_{uv}$. As for $\mathcal{L}_{uu}$ and $\mathcal{L}_{vv}$, even though the computation involves the summation over the interactions between a user/item and all the groups (i.e. $\sum_{k=1}^Kexp({\mathbf{g}_k^u}^T\mathbf{z}^{u_i})$), the number of groups are much fewer than that of users/items, and hence the summation does not incur high computational overhead. In the proposed DBRec model, we jointly define the loss function, so that it can be optimized efficiently by backpropagation. Taking into account the loss of dual-bridging collaborative filtering, hierarchy modeling and group learning, the final objective function of DBRec can be defined as follows, 
\begin{equation}
\label{eq:objective}
\mathcal{L} = \mathcal{L}_{uv}+\alpha(\mathcal{L}_{uu}+\mathcal{L}_{vv}+\mathcal{L}_u+\mathcal{L}_v)
\end{equation}
where $\alpha$ controls the tradeoff between the loss of dual-bridging collaborative filtering and that of hierarchy modeling and group learning. The training of the proposed model can be decomposed into two parts. In the part of collaborative filtering and hierarchy modeling, user and item embeddings are updated to fit user-item interaction data and preserve group hierarchies. While in the part of group learning, group embeddings are updated to discover representative group information.

The model parameters include user/item representations and parameters of neural networks as shown in Fig.\ref{fig:lgd}, Fig.\ref{fig:hierarchy} and Fig.\ref{fig:network}. We update the object function with Adam optimizer, which is a variant of Stochastic Gradient Descent with a dynamically tuned learning rate and updates parameters every step along the gradient direction with the following protocol:
\begin{equation}
\boldsymbol{\theta}^{t}\leftarrow\boldsymbol{\theta}^{t-1}-lr\frac{\partial\mathcal{L}}{\partial\boldsymbol{\theta}}
\end{equation}
where $lr$ is the learning rate, and $\boldsymbol{\theta}$ are the model parameters and $\frac{\partial\mathcal{L}}{\partial\boldsymbol{\theta}}$ are the partial derivatives of the objective function with respect to the model parameters, and they can be automatically computed with typical deep learning libraries.

\section{Experiments}
\subsection{Experimental Settings}
\subsubsection{Datasets}
\begin{table}
\caption{Statistics of the datasets.}
\label{tb:data}
\centering
\begin{tabular}{l|cccc}
\hline
\textbf{Dataset} & \textbf{\#user} & \textbf{\#item} & \textbf{\#ratings} &\textbf{Sparsity}\\
\hline
\textit{Amazon} & 4016 & 7199 & 32190 & 99.89\% \\
\hline
\textit{ML1M} & 6040 & 3706 & 1000209 & 95.53\% \\
\hline
\textit{Gowalla} & 78477 & 528968 & 5380147 & 99.98\%\\
\hline
\end{tabular}
\end{table}

To validate the effectiveness of the proposed model, we conduct experiments on three publicly datasets, namely Amazon\footnotemark[2], MovieLens-1M (ML1M)\footnotemark[1] and Gowalla\footnotemark[3]. For the amazon dataset, we select the items of musical instruments for evaluation. To transform the explicit ratings of Amazon and ML1M into implicit feedback, ratings higher than 3 are regarded as positive feedback, and the others are regarded as missing values. As for Gowalla, for each user the check-in locations are viewed as positive items, while the other locations are treated as missing items. Inspired by previous works \cite{Liang2016modeling, Lian2014geomf}, we filter out users with fewer than 5 positive items and the items with fewer than 2 users. The statistics of the datasets is shown in Table.\ref{tb:data}.

\footnotetext[1]{https://grouplens.org/datasets/movielens/}
\footnotetext[2]{http://jmcauley.ucsd.edu/data/amazon/} 
\footnotetext[3]{http://snap.stanford.edu/data/loc-gowalla.html}

\subsubsection{Setup}
To evaluate the proposed model, we randomly split the datasets into training set (70\%), validation set (10\%) and testing set (20\%). In the training process, for each positive item, we randomly sample 5 items as the negative samples. In the testing process, as it is time-consuming to rank all the items for every user at each time, hence for each $(u_i,v_j)$ pair in the testing set, we mix the testing item with 99 random items, and rank the testing item along with the 99 items for the related user. We measure the recommendation performance with the commonly used Hit Ratio (HR) and Normalized Discounted Cumulative Gain (NDCG), as shown follows,
\begin{equation}
\begin{split}
HR&=\frac{\#hits}{\#test}\\
NDCG&=\frac{1}{\#test}\sum_{i=1}^{\#test}\frac{1}{log_2(p_i+1)}
\end{split}
\end{equation}
where $\#hits$ is the number of testing item that appears in the the recommendation list of the related user and $\#test$ is the total number of $(u_i,v_j)$ pair in the testing set. $p_i$ is the position of the testing item in the recommendation list for the $i$-th hit. HR measures whether the testing item is in the recommendation list, while NDCG assigns higher score to the testing item with higher position. In this paper, we truncate the ranking list at $k\in[1,2,\cdots,10]$ for both metrics.
\subsubsection{Baselines}
The baselines employed for performance comparison are list as follows, 
\begin{itemize}
\item NeuMF\cite{he2017neural}. It combines generalized MF and Multi-Layer Perception (MLP) for modeling user-item latent structures. 

\item ConvMF\cite{he2018outer}. It uses outer product to transform latent vectors of a user-item pair into two-dimensional map, and applies convolutional and pooling layers to model deep user-item interactions. 

\item CMN \cite{ebesu2018collaborative}. It identify similar neighboring users with an attention mechanism based on the specific user-item pair, and jointly exploits the neighborhood state and user-item interactions to derive recommendation. 

\item DELF\cite{cheng2018delf}. It proposes an attention mechanism to aggregate an additional embedding for each user/item, and then further introduce a neural network architecture to incorporate dual embeddings for recommendation. 
\end{itemize}

For fair comparison, baselines in this paper mainly consist of the state-of-the-art recommendation models based on user-item interactions, and the hyper-parameters of these baselines are set according to the original works. Notice that even though CMN exploits neighborhood information for recommendation, the neighboring users are selected mainly based on their rating behaviours. Beside those baselines, we also study various variants of the proposed model, list as follows,
\begin{itemize}
\item DBRec-o, a variant of the proposed model that excludes user and item group learning, and it mainly relies on user-item interactions for learning user preferences over the items. 

\item DBRec-u, a variant of the proposed model that excludes item group learning, and it jointly learns user group and model user-group hierarchy, and interacts user group information with item for boosting recommendation.

\item DBRec-i, a variant of the proposed model that excludes user group learning, and it jointly learns item group and model item-group hierarchy, and interacts item group information with users for boosting recommendation.
\end{itemize}

\subsubsection{Implementation}
We implement DBRec based on Tensorflow, and the sources are made publicly available\footnotemark[3] to facilitate community research. We set the batch size to 256 and the learning rate to 0.0001.The dimensions of latent factor is 128. We employ two hidden layers for the neural networks illustrated in Fig.\ref{fig:network} and Fig.\ref{fig:hierarchy}, and the hidden units of the respective layer are [64,16] for dual-bridging collaborative filtering, and [64,128] for hierarchy modeling. The model parameters are fine tuned on the validation set. We set the number of user/item group to 5, and the trade-off weight (i.e. $\alpha$ in Eq.(\ref{eq:objective})) for group learning and hierarchy modeling to 0.01. We analyze the sensitivity of these two hyper-parameters later in this section.

\footnotetext[3]{https://github.com/uqjwen/DBRec}

\paragraph{pre-training} The objective function is non-convex, and the learning method can be easily trapped in local optimums. In our work, we pre-train DBRec with NeuMF to obtain initial user and item embeddings in DBRec. As for the group embeddings, we perform clustering on the pre-trained user and item embeddings, and use the cluster centroids to initialized the group embeddings. Since users/items and their corresponding groups do not share the same latent space, we employ a dimensionality reduction method (i.e. t-SNE \cite{maaten2008visualizing}) to reduce the centroids of $d$-dimension to that of $d_g$ dimensions. 

\subsection{Experiments}
\subsubsection{Recommendation comparisons}

\begin{figure}
  \centering
  \subfigure[Amazon]{
  \includegraphics[width=0.7\linewidth]{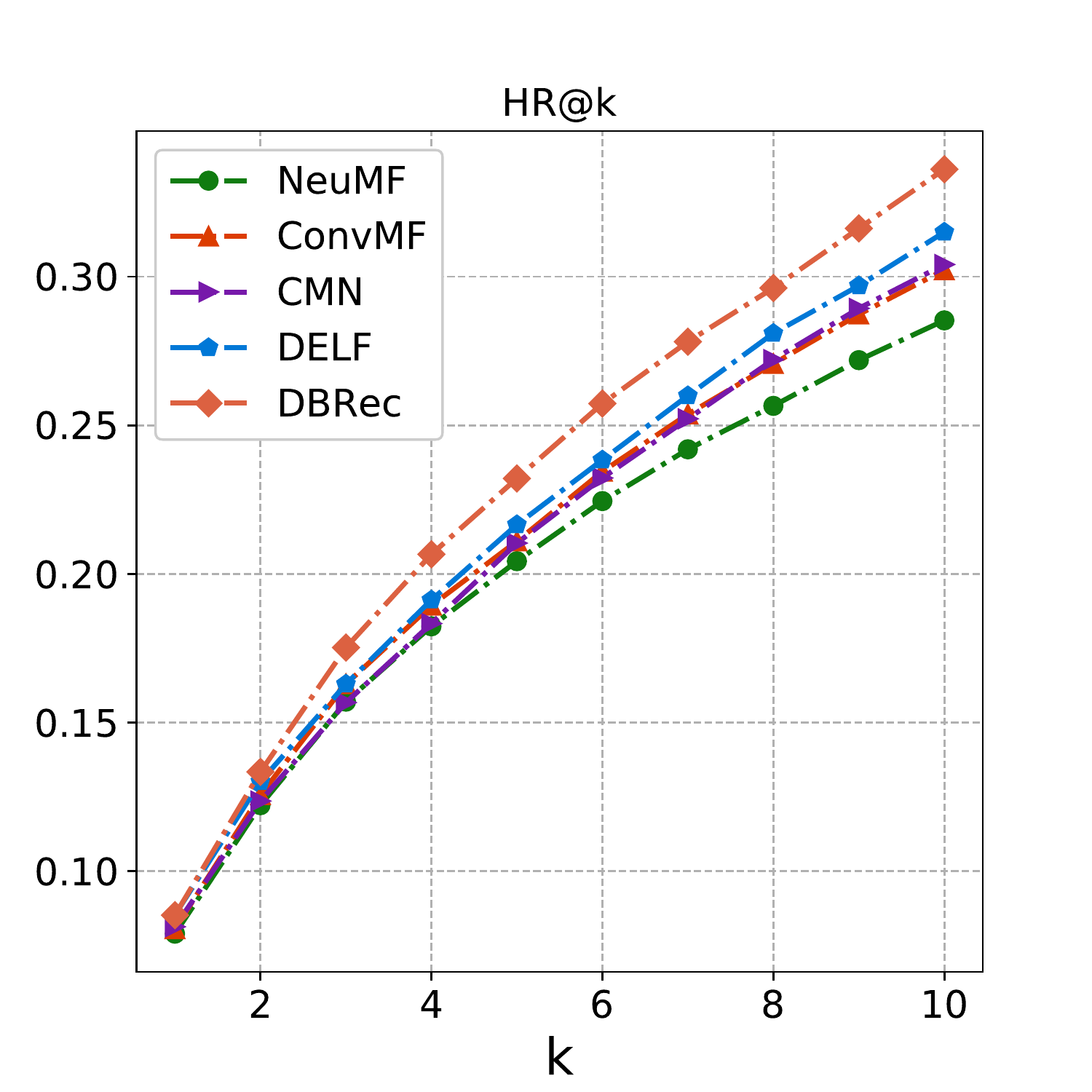}}
  \subfigure[ML1M]{
    \includegraphics[width=0.7\linewidth]{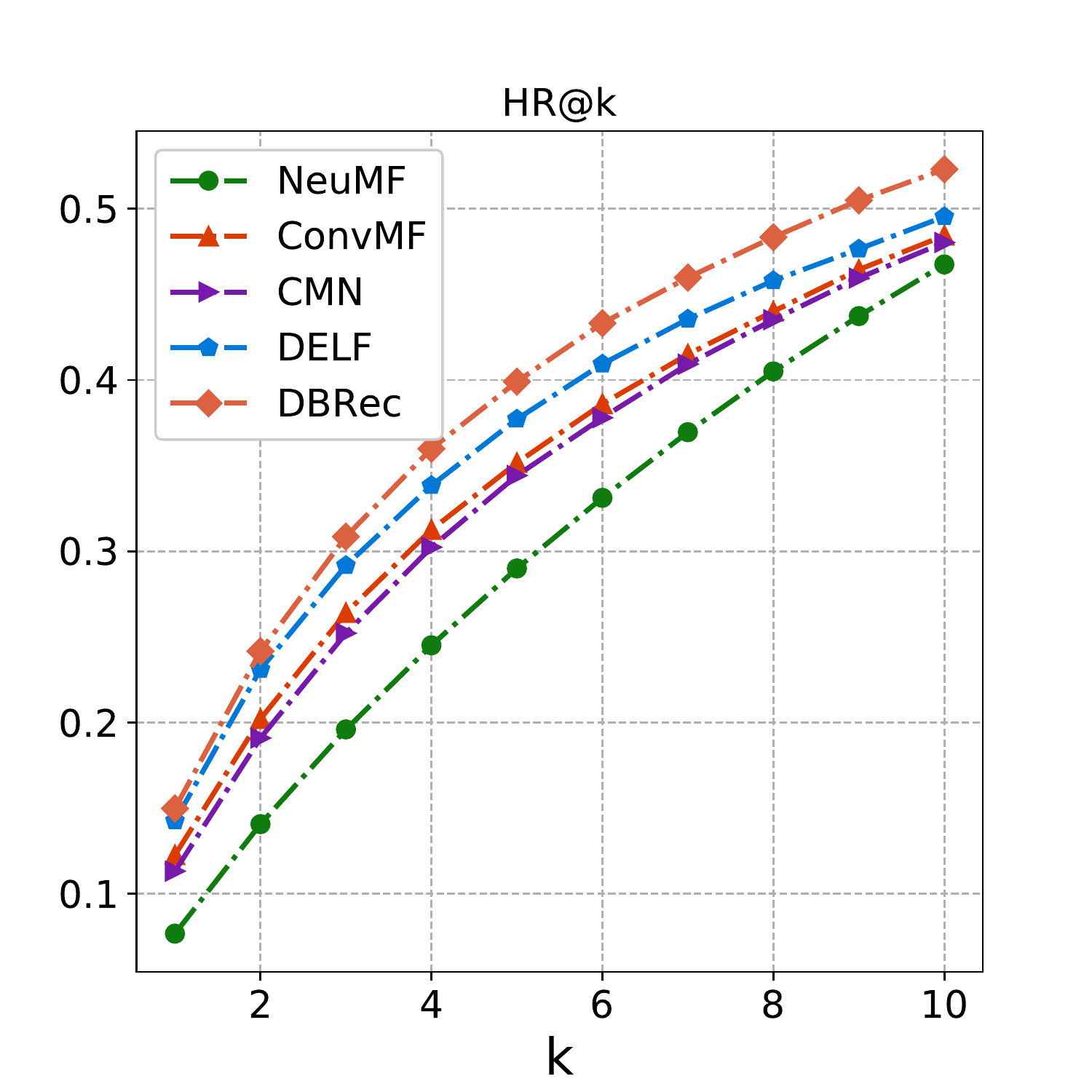}}
  \subfigure[Gowalla]{
    \includegraphics[width=0.7\linewidth]{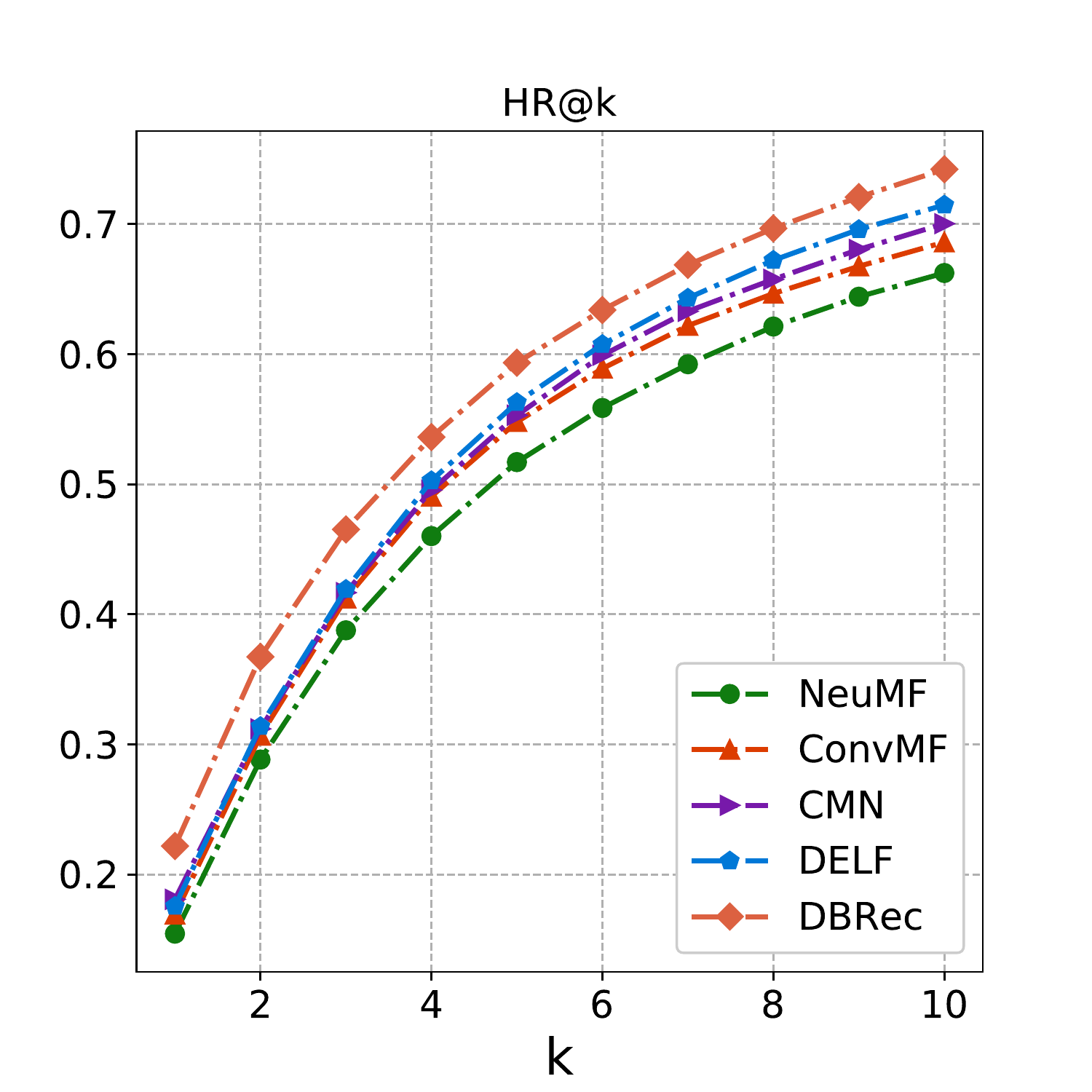}}
  
   
  \caption{HR of different models across the datasets}
  \label{fig:hr} 
\end{figure}

\begin{figure}
  \centering
  \subfigure[Amazon]{%
    \includegraphics[width=.7\linewidth]{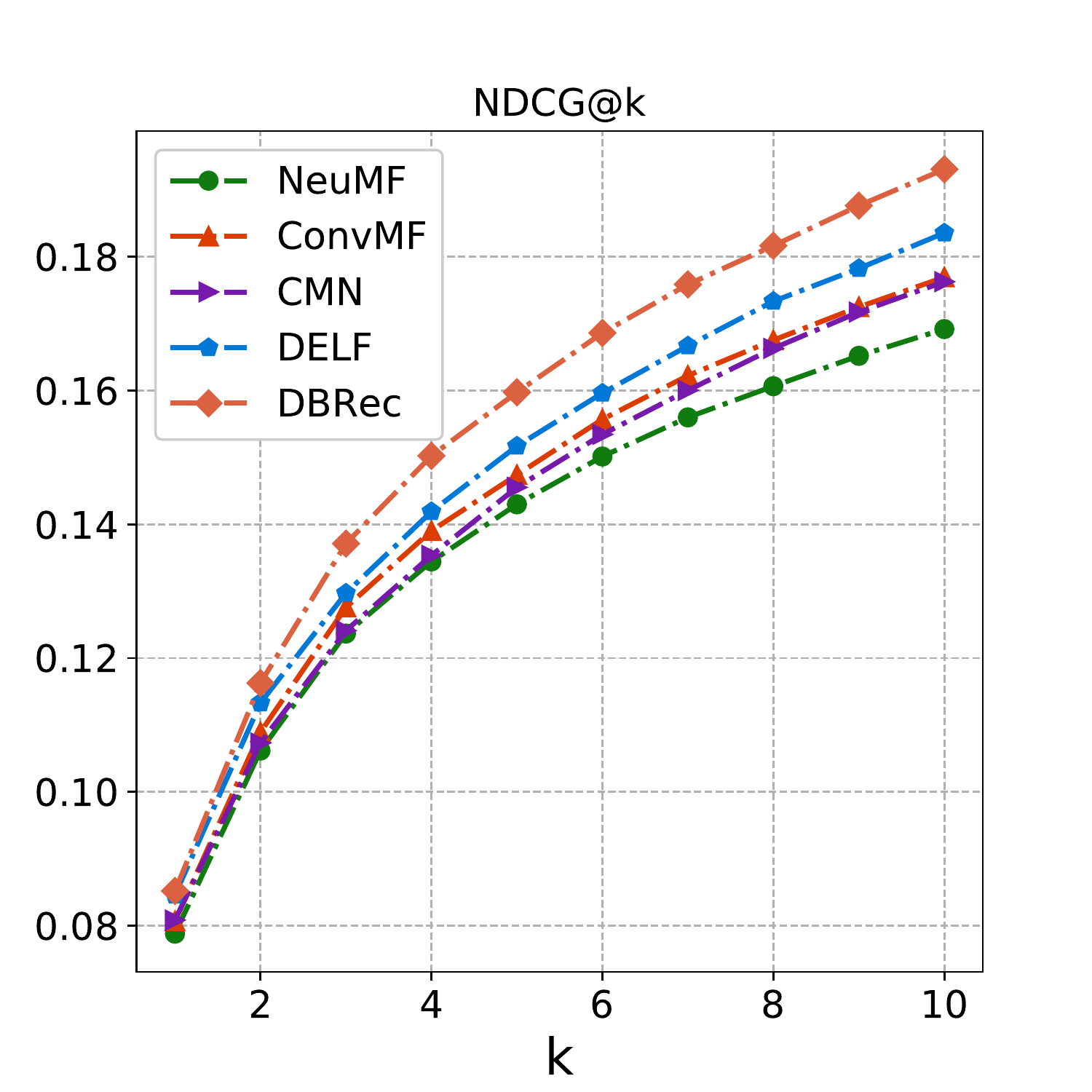}}
  \subfigure[ML1M]{%
    \includegraphics[width=.7\linewidth]{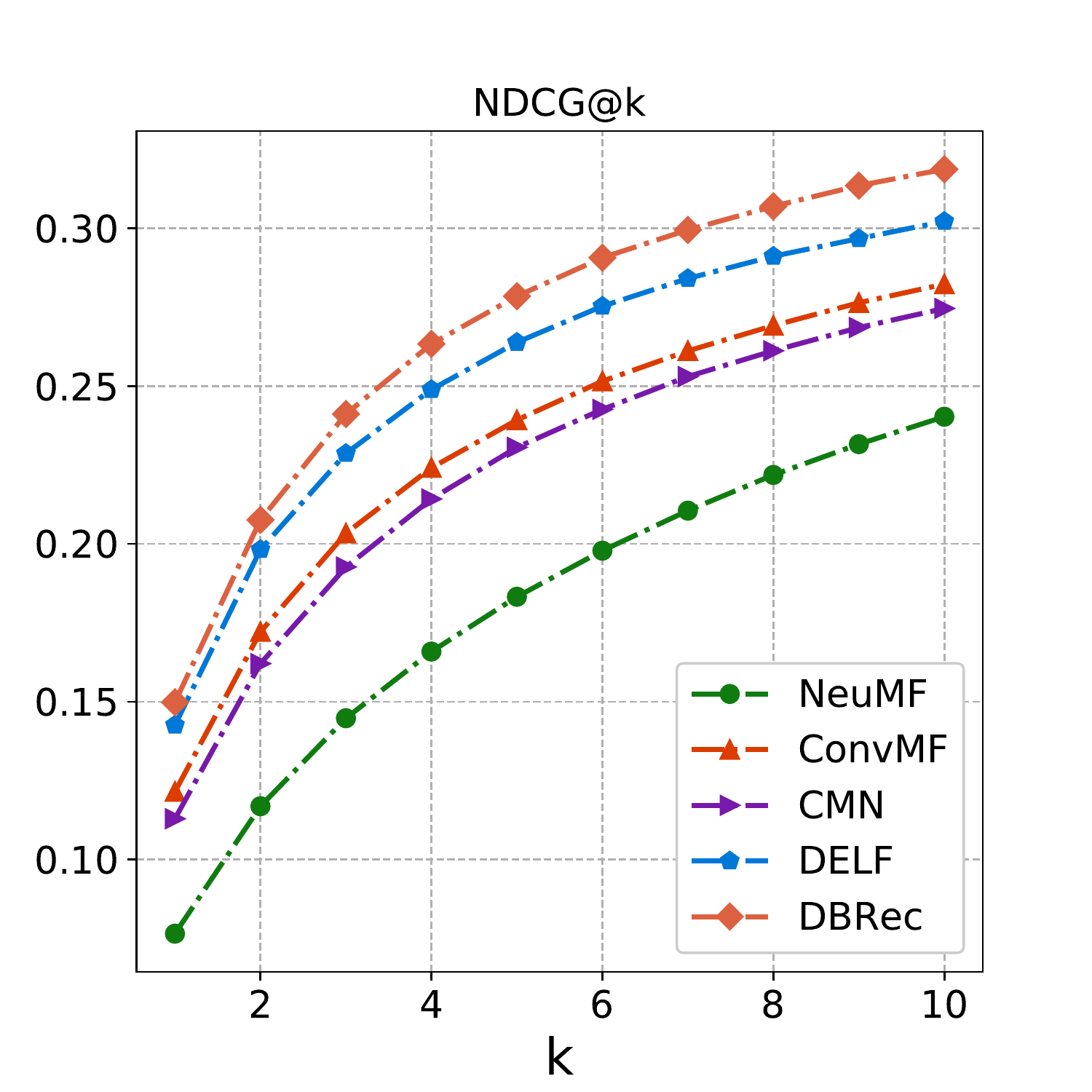}}
  \subfigure[Gowalla]{%
    \includegraphics[width=.7\linewidth]{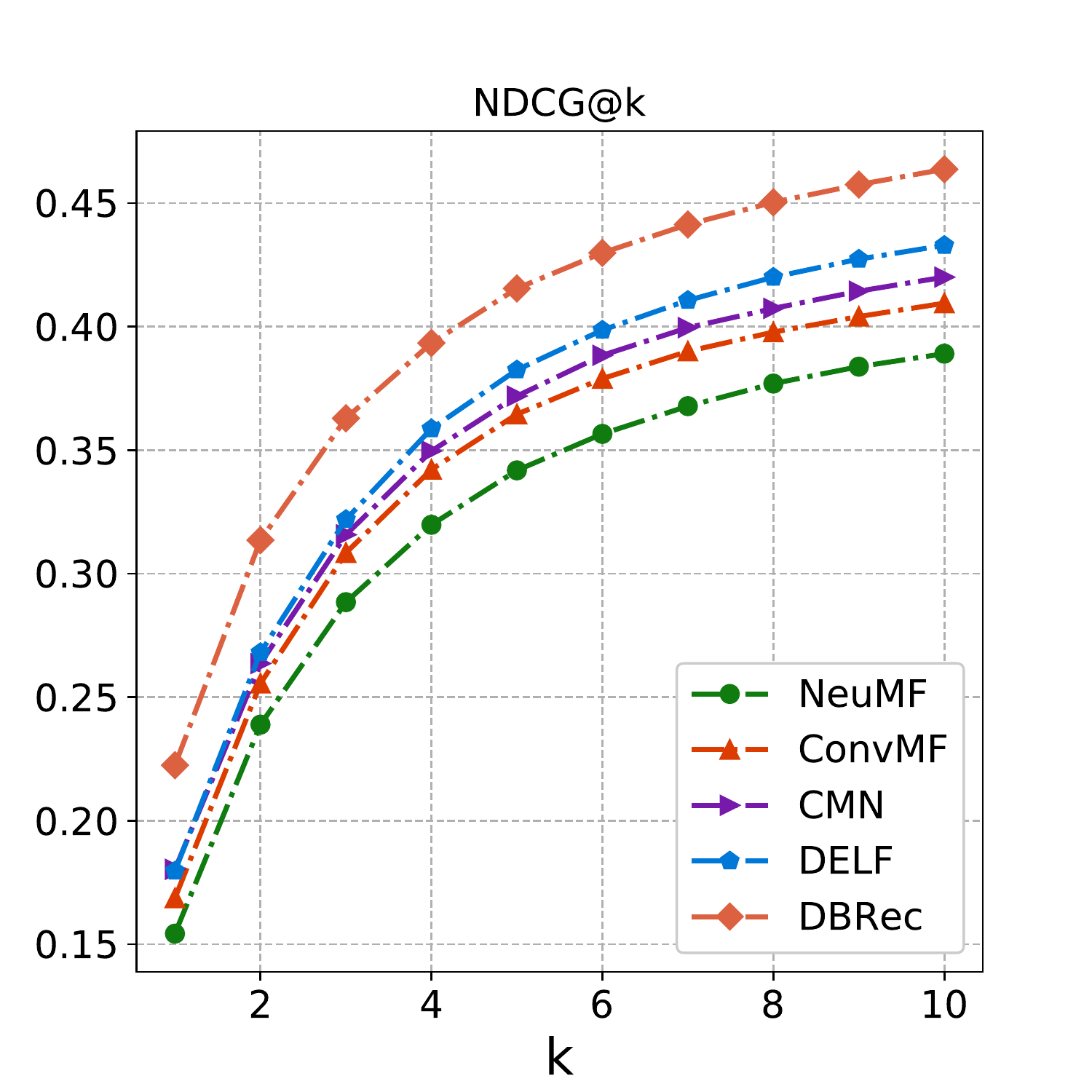}}
   
  \caption{NDCG of different models across the datasets.}
  \label{fig:ndcg} 
\end{figure}

We compare the proposed model, DBRec, with the state-of-the-art recommendation models in terms of top-K recommendation, and the results are presented in Fig.\ref{fig:hr} and Fig.\ref{fig:ndcg}. From the figures, we have the following observations. First, DELF, ConvMF and CMN outperform NeuMF across different datasets and metrics, which is consistent with previous works \cite{he2018outer, ebesu2018collaborative, cheng2018delf}. The reason is that NeuMF mainly relies user-item interactions for exploiting user preferences, and it suffers from data sparseness due to the sparsity nature of the rating matrix. Second, the performance differences between CMN and ConvMF vary across different datasets. For example, ConvMF achieves marginally better performance than CMN on ML1M, while CMN outperforms ConvMF on Gowalla. This is because they leverage different aspects of interaction data for boosting performance. Specifically, ConvMF utilizes outer product to exploit dimension-wise correlations for recommendation, while CMN exploits informative neighboring users for bridging each user-item pair. Therefore, the performance differences between these two models are quite data-dependent. Third, the figures show that DELF is superior to CMN, and one possible explanation might be that DELF aggregates for each user/item an additional embedding based on the interacted items/users, and introduce a neural network to incorporate the dual embeddings for recommendation. In addition, DELF can identify the most informative context users/items for collaborative filtering, due to the proposed effective attention mechanism.

Finally, the proposed model outperforms the baselines by a large margin across the datasets for both HR and NDCG. Specifically, in terms of top-10 recommendation, DBRec outperforms the best baseline, DELF, with a relative improvement of 6.71\%(HR@10) and 5.19\%(NDCG@10) on Amazon, 5.59\%(HR@10) and 5.46\%(NDCG@10) on ML1M, and 3.84\%(HR@10) and 7.11\%(NDCG@10) on Gowalla, respectively. The reason behinds the advantage of DBRec over the baselines is that, we collaboratively discover latent group information for collaborative filtering, and the incorporation of group information can capture user preferences (item characteristics) at different levels of granularity and mitigate data sparseness.

\subsubsection{Efficacy of dual-bridging architecture}
\begin{table*}
\caption{Performance of variants of DBRec.}
\label{tb:variant}
\setlength{\tabcolsep}{7mm}
\begin{tabular}{ll|llll}
\hline
\textbf{Datasets} & \textbf{Metrics} & \textbf{DBRec-o} & \textbf{DBRec-i} & \textbf{DBRec-u} & \textbf{DBRec}\\
\hline
\hline
\multirow{4}*{Amazon} & HR@5&0.20429&0.22606&0.213&0.23212\\
~& HR@10&0.28529&0.324&0.31281&0.33613\\
~& NDCG@5&0.143&0.15832&0.14688&0.15975\\
~& NDCG@10&0.1692&0.19001&0.17924&0.19309\\
\hline
\multirow{4}*{ML1M} & HR@5&0.28992&0.37152&0.36942&0.39906\\
~& HR@10&0.46752&0.51233&0.50847&0.52311\\
~& NDCG@5&0.18326&0.25381&0.25183&0.27845\\
~& NDCG@10&0.24028&0.29936&0.29685&0.31865\\
\hline
\multirow{4}*{Gowalla} & HR@5&0.56284&0.5757&0.58372&0.59342\\
~& HR@10&0.71583&0.72899&0.73358&0.74211\\
~& NDCG@5&0.3825&0.39376&0.39768&0.41547\\
~& NDCG@10&0.43291&0.44354&0.44637&0.46371\\
\hline
\end{tabular}
\end{table*}

In this subsection, we study the performance of various variants of the proposed model, to validate the effectiveness of the proposed dual-bridging architecture. As shown in Table.\ref{tb:variant}, without latent group information, DBRec degrades to the version (DBRec-o) that is mainly based on user-item interactions, and is vulnerable to data density. However, by discovering latent groups and incorporating group information for collaborative filtering, DBRec-i and DBRec-u significantly outperform DBRec-o, demonstrating the benefit of discovering latent groups for bridging similar users/items.

DBRec combines DBRec-i and DBRec-u, and results in a dual-bridging architecture, which models user preferences and item characteristics at different levels of granularity. General group representations can alleviate the problem of data sparseness by bridging similar users/items, and achieves better recommendation performance than otherwise single-bridging model (e.g. DBRec-i, DBRec-u). 

\subsubsection{Sensitivity analysis}
\begin{figure}
  \centering
  \subfigure[]{%
    \includegraphics[width=.45\linewidth]{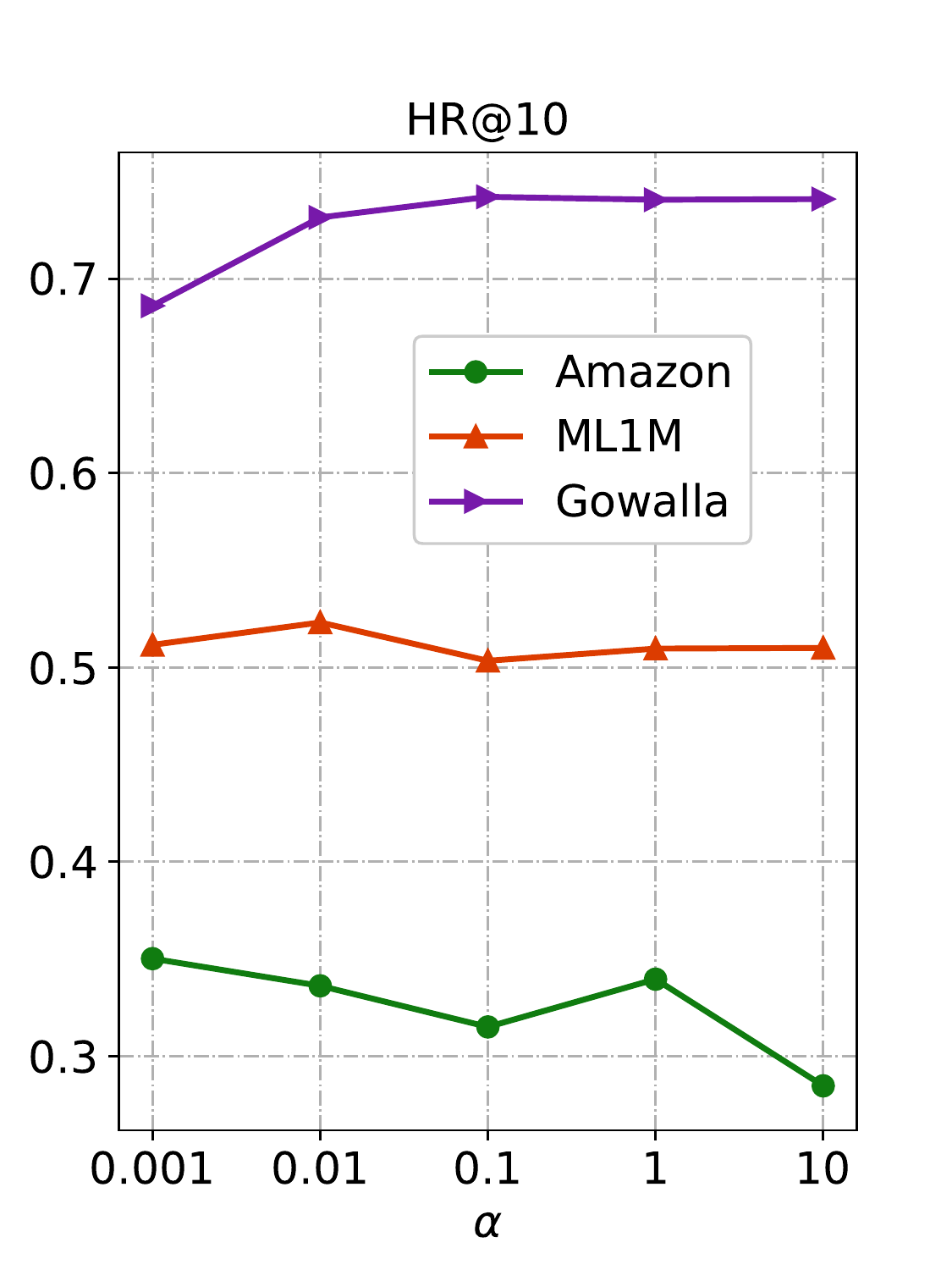}}
  \subfigure[]{%
    \includegraphics[width=.45\linewidth]{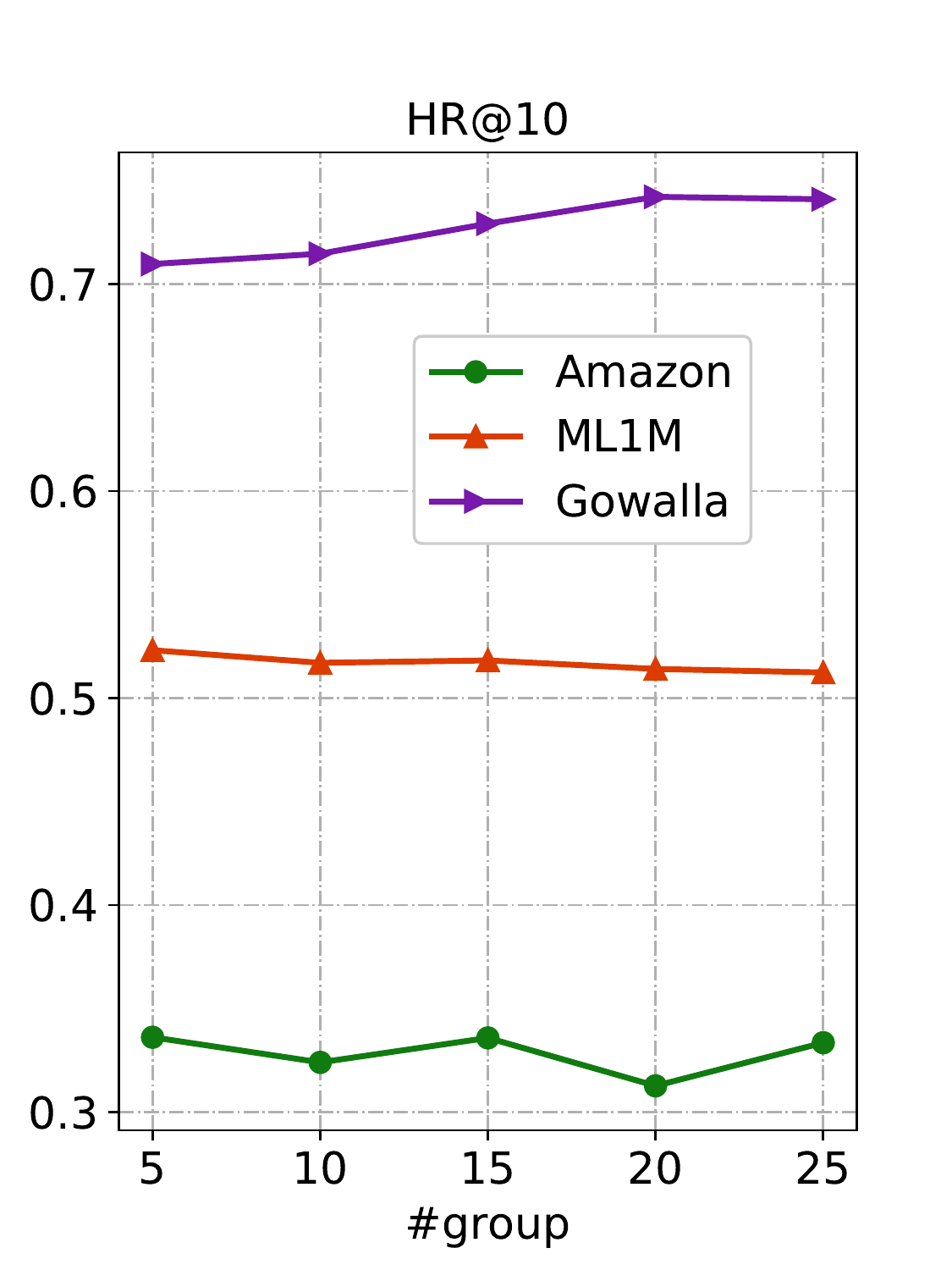}}
  \subfigure[]{%
    \includegraphics[width=.45\linewidth]{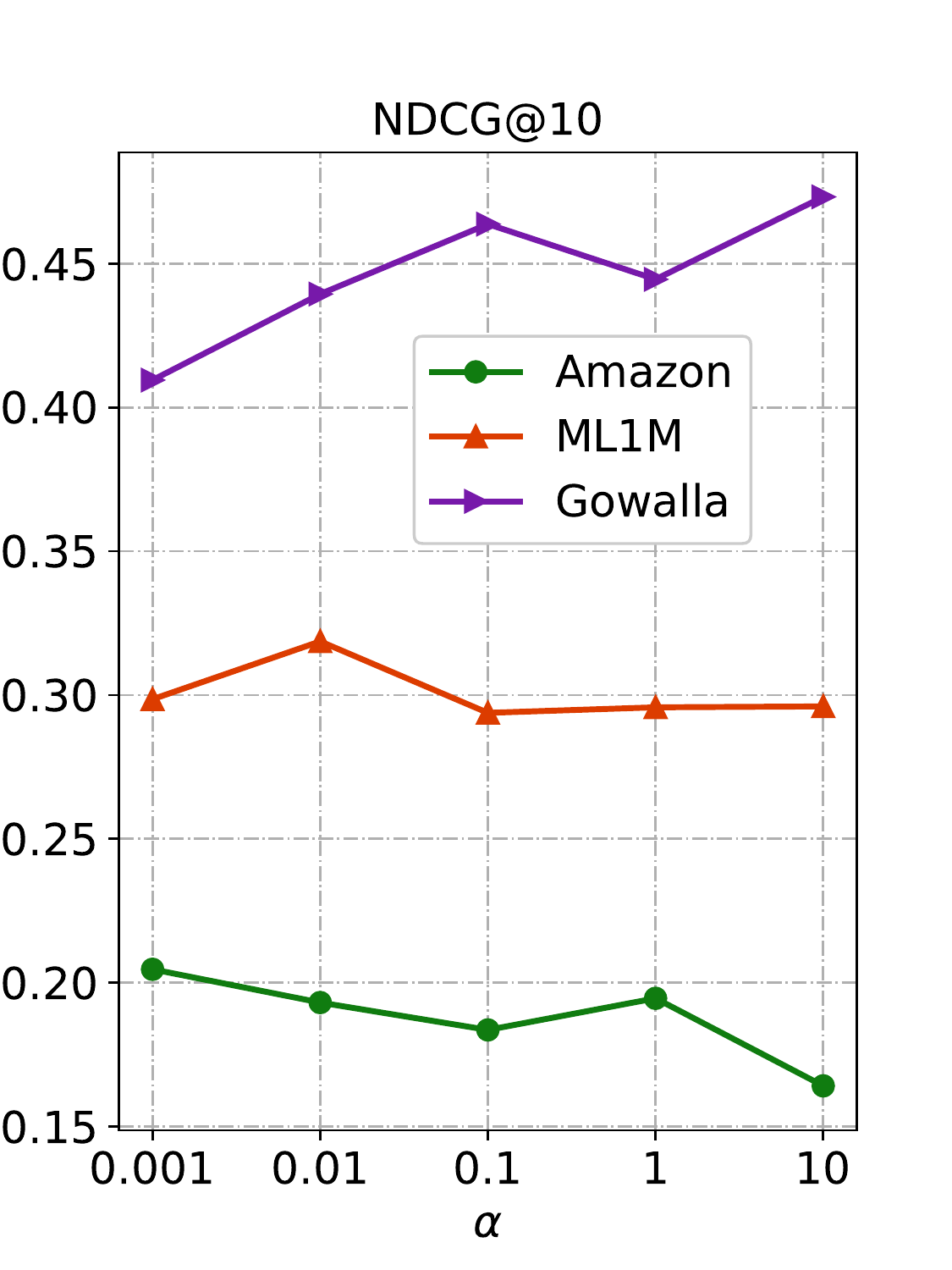}}
  \subfigure[]{%
    \includegraphics[width=.45\linewidth]{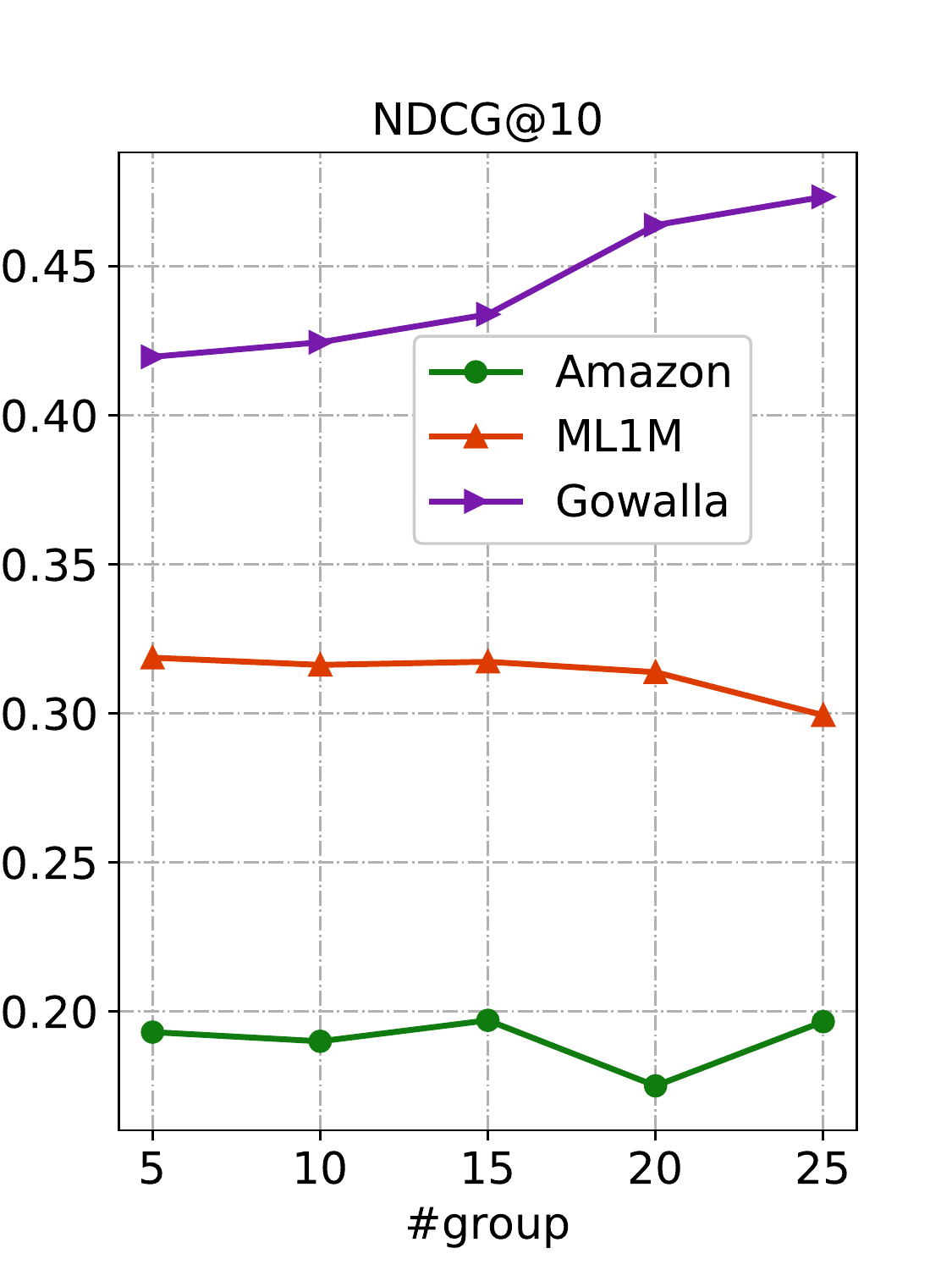}}
  \caption{Recommendation performance of the proposed model with respect to different hyper-parameters.}
  \label{fig:paras} 
\end{figure}

In this subsection, we investigate the robustness of the proposed model with respect to different hyper-parameters. Especially, we study the performance of DBRec under different pre-specified user/item group numbers, and the weight $\alpha$ (Eq.(\ref{eq:objective})) for group learning and hierarchy modeling. In this experiment, $\alpha$ is varied amongst [0.001,0.01,0.1,1,10], and the group number [5,10,15,20,25]. As illustrated in Fig.\ref{fig:paras}, we present the performance in terms of HR@10 and NDCG@10 across the datasets. One can see from the figures that, DBRec experiences a rapid performance degradation with $\alpha=10$ and $\#group=20$ on Amazon. However, the recommendation performance is relatively stable under a wide-range choice of the hyper-parameters. As shown in Fig.\ref{fig:paras}(b) and Fig.\ref{fig:paras}(d), Gowalla presents a steady performance improvement with more group number, probably because larger group number is required to preserve group information due to the size of the dataset. HR@10 and NDCG@10 usually show the similar trends across the datasets except for Gowalla, which experiences a sudden drop on NDCG@10 with $\alpha=1$, but the expected phenomenon is not observed on HR@10 with $\alpha=1$, as shown in Fig.\ref{fig:paras}(a) and Fig.\ref{fig:paras}(c). The reason for the slump of NDCG on Gowalla is that, the testing items are ranked lower at the bottom of the recommendation list. Overall, this experiment demonstrates that the proposed model is invulnerable to the group number. Even though the users/items are clustered into different granularities, the group information can still boost recommendation performance as long as discriminate information provided by the user/item groups can be preserved in the clustering process.

\subsubsection{Visualization}
\begin{figure*}
  \centering
  \subfigure[user/Amazon]{%
    \includegraphics[width=.3\linewidth]{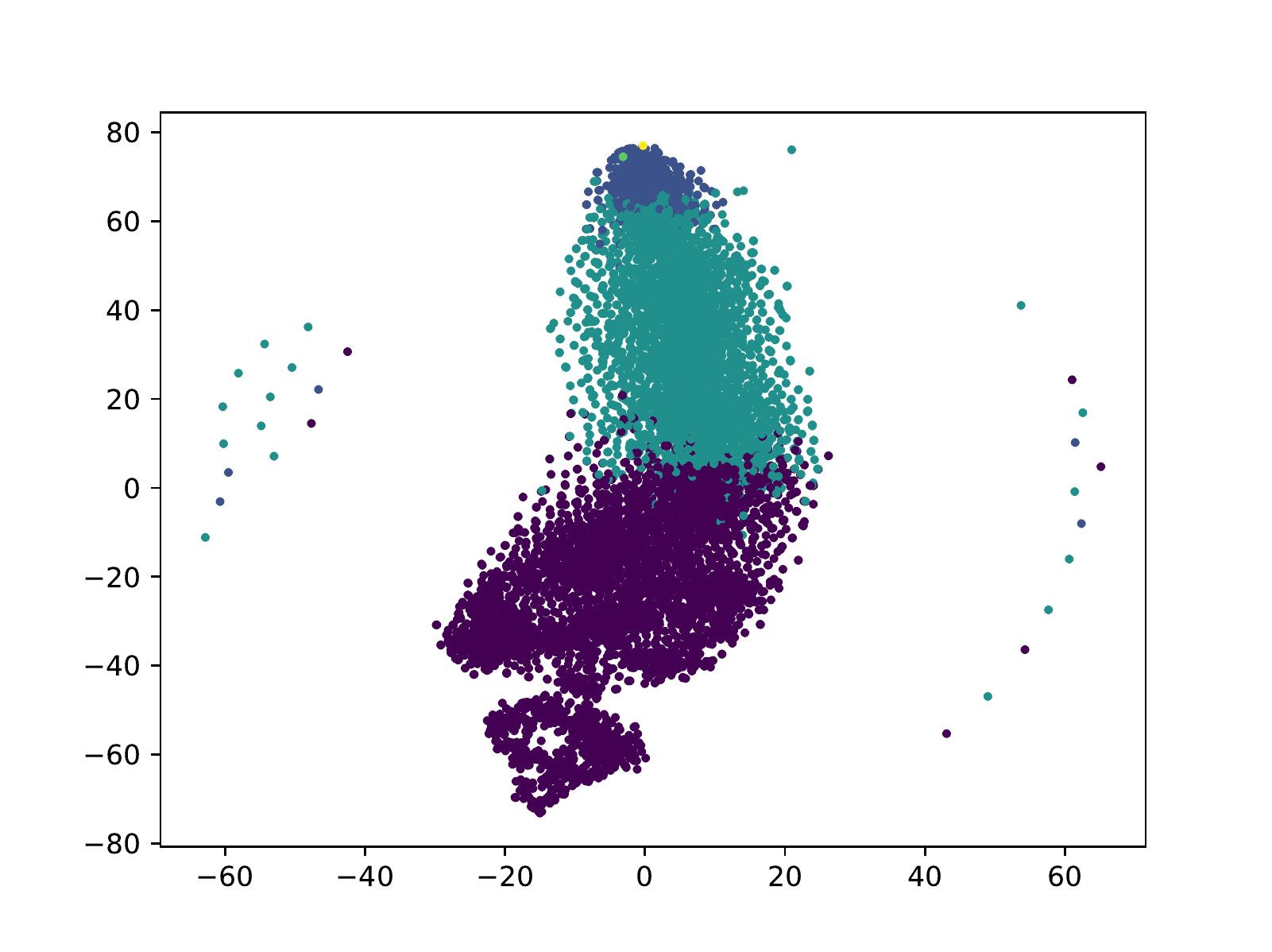}}
  \subfigure[user/ML1M]{%
    \includegraphics[width=.3\linewidth]{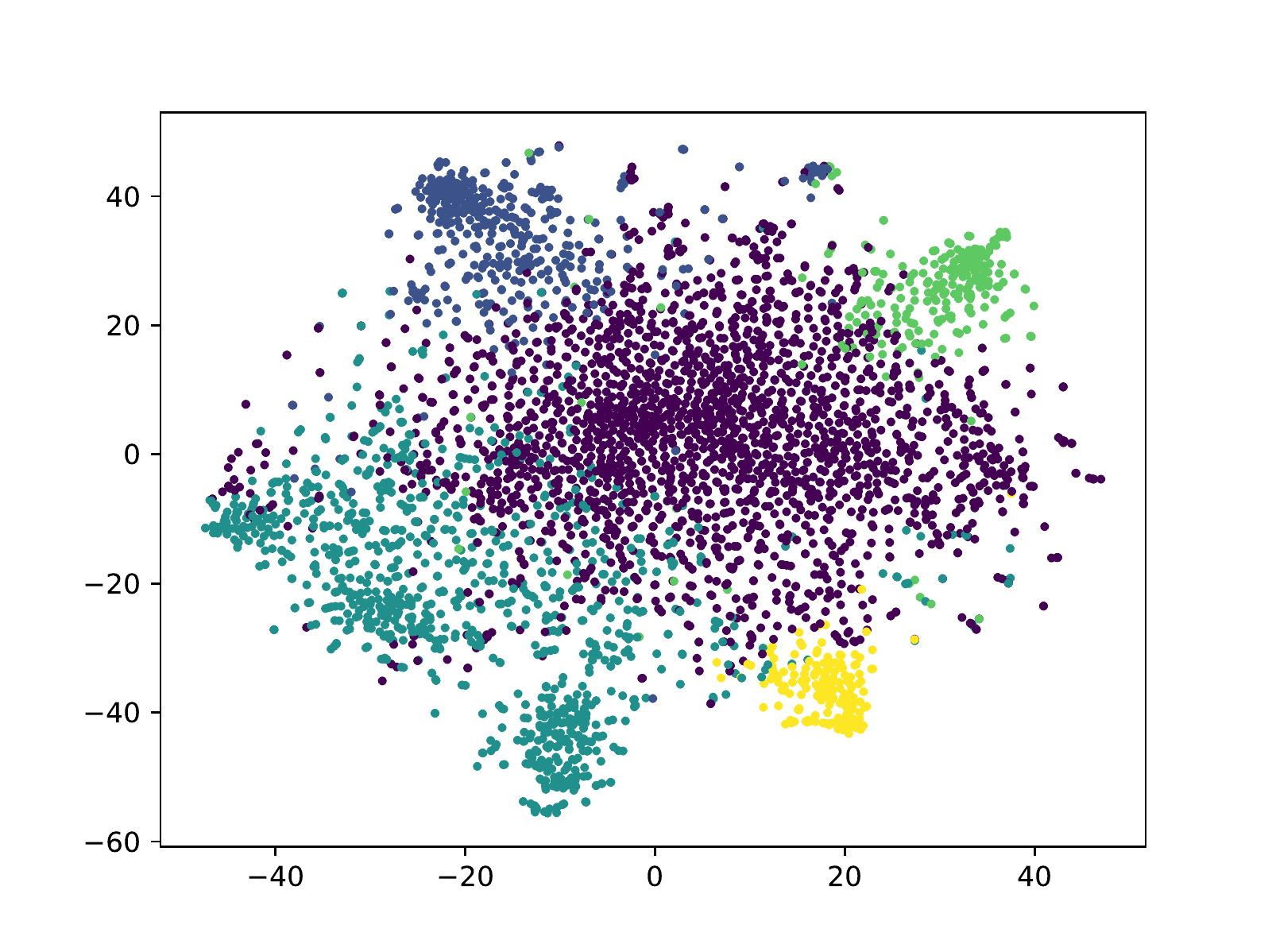}}
  \subfigure[user/Gowalla]{%
    \includegraphics[width=.3\linewidth]{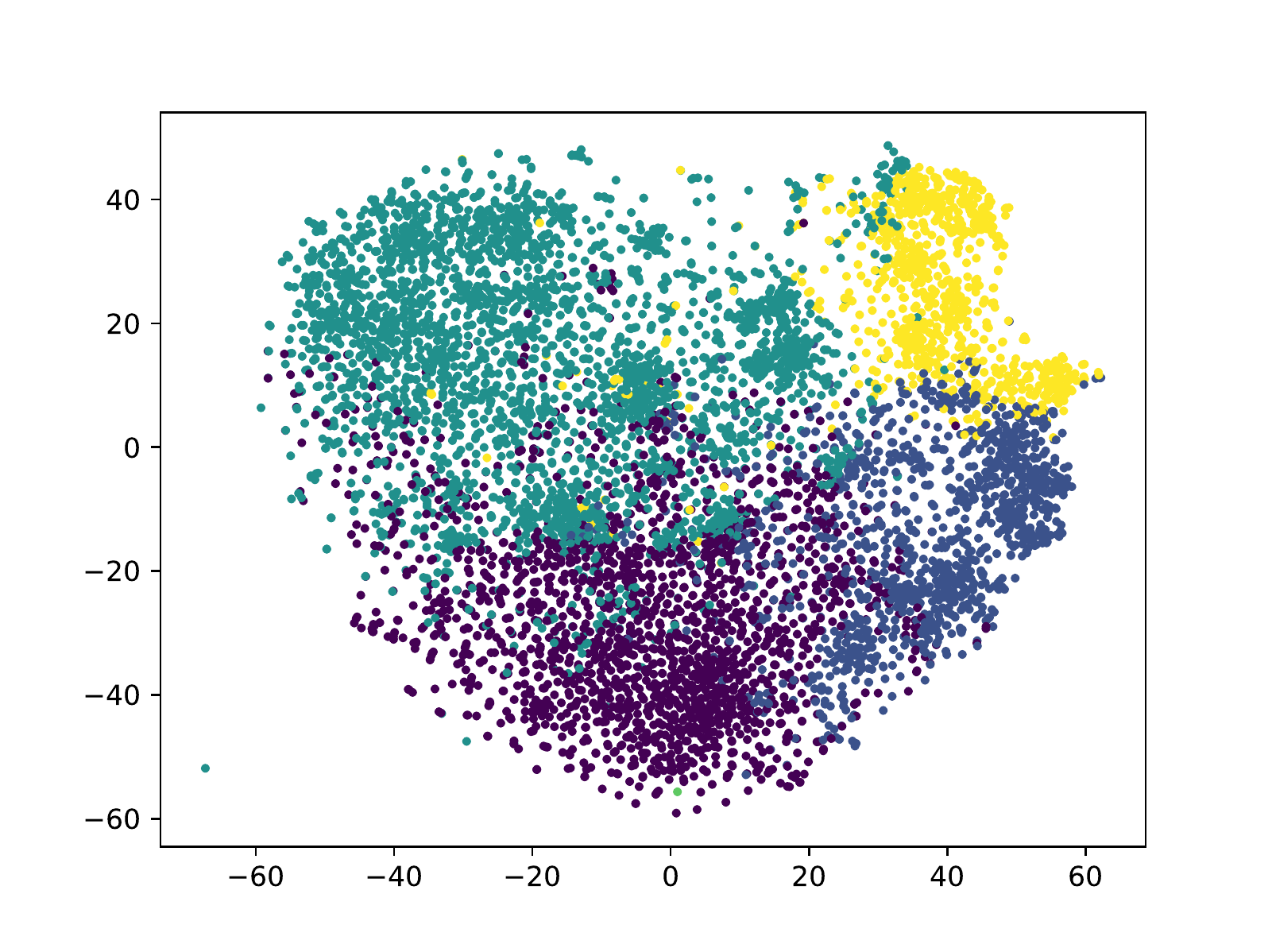}}
  \caption{visualizing latent user groups across the datasets}
  \label{fig:visua} 
\end{figure*}

To examine whether DBRec can discover latent groups, we visualize the learned user representations with t-SNE \cite{maaten2008visualizing}, and plot users in the same group with the same color. As shown in Fig.\ref{fig:visua}, several user/item groups can be clearly identified. This experiment demonstrates that DBRec can effectively discover latent groups, and visually interpret the rationale behind the dual-bridging architecture in DBRec.

\section{Conclusion}
In this paper, we propose a novel dual-bridging recommendation model. The advantage of the proposed model is that it can discover latent group information for bridging similar users/items, and boost recommendation performance. We unified collaborative filtering, group learning and hierarchy modeling into a framework, so that all the sub-components can be jointly optimized and compensate each other for boosting recommendation. Extensive experiments on two real datasets demonstrate the advantage of DBRec in terms of recommendation performance, and robustness in terms of parameters sensitivity.

The proposed model in this paper mainly relies on user-item interaction data, and discover extra latent group information for boosting recommendation performance. In real recommendation scenarios, many attribute data are explicitly available, and integrating heterogeneous modalities of implicit and explicit information for recommendation is an interesting research topic in the future. 

\section{Acknowledgments}
This publication was made possible by Opinion Analysis grant 018493 from Australian Research Council Fund and NPRP grant NPRP10-0208-170408 from the Qatar National Research Fund (a member of Qatar Foundation).

\bibliographystyle{ACM-Reference-Format}
\bibliography{dbrec}

\end{document}